# Thermodynamic Limit for Linear Harmonic Oscillator Resonance Frequency Measurement

*Mingkang Wang[1,2] and Vladimir Aksyuk[1]*
[1]Microsystems and Nanotechnology Division, National Institute of Standards and Technology, Gaithersburg, MD 20899 USA
[2]Institute for Research in Electronics and Applied Physics, University of Maryland, College Park, MD 20742, USA

**Thermodynamic fluctuations in mechanical resonators cause uncertainty in their frequency measurement, fundamentally limiting performance of frequency-based sensors. Recently, integrating nanophotonic motion readout with micro- and nano-mechanical resonators allowed practical chip-scale sensors to routinely operate near this limit in high-bandwidth measurements. However, the exact and general expressions for either thermodynamic frequency measurement uncertainty or efficient, real-time frequency estimators are not well established, particularly for fast and weakly-driven resonators. Here, we derive, and numerically validate, the Cramer-Rao lower bound (CRLB) and an efficient maximum-likelihood estimator for the frequency of a classical linear harmonic oscillator subject to thermodynamic fluctuations. For a fluctuating oscillator without external drive, the frequency Allan deviation calculated from simulated resonator motion data agrees with the derived CRLB $\sigma_f = \frac{1}{2\pi}\sqrt{\frac{\Gamma}{2\tau}}$ for averaging times $\tau$ below, as well as above, the relaxation time $\frac{1}{\Gamma}$. The CRLB approach is general and can be extended to driven resonators, non-negligible motion detection imprecision, as well as backaction from a continuous linear quantum measurement.**

Nano-electro-mechanical systems (NEMS) have drawn intensive interest from both the applied and the fundamental perspectives in the last decade. They have proved to be not only qualified platforms to study fundamental science such as quantum physics [1]–[4] and nonlinear dynamics [5], [6] but also key components for various detection schemes, such as magnetic resonance microscopy [7], [8], mass spectrometry [9]–[11] and well-known atomic force microscopy [12]. The miniaturization of the NEMS, down to atomically-defined structures such as carbon nanotubes, leads to extraordinary sensitivity for force [13], mass [14], as well as quantities such as charge [15] and magnetic torque [16]. Typically, the unknown parameter is converted into a change of the resonance frequency, which is measured via the resonator motion. Thanks to the rapid development of low-noise optical transduction techniques, e.g. cavity optomechanical readout [17], [18], the motion signal and its thermal fluctuations are well resolved above the detection noise in ever more broadband measurements. These developments demand a quantitative understanding of the fundamental thermodynamic limits on the frequency measurement precision across a range of measurement bandwidths and drive strengths. Besides, a computationally-fast and statistically efficient frequency estimator – an algorithm for converting motion records into frequencies with imprecisions not exceeding their fundamental limits – is also needed.

The problem of resonator frequency measurement and stabilization in the context of micro- and nanomechanical systems continues to receive researcher attention. The frequency stabilization is implemented by various approaches, such as structure engineering [19] or feedback control [20]–[22] in the linear regime, and using model coupling [6] or zero-dispersion point [5] in the nonlinear regime. Most work is generally focused on strongly externally driven oscillators, e.g. in one notable recent report frequency uncertainty was improved with lower intrinsic quality factor resonators, which could be more strongly driven before the onset of nonlinearity [23]. However, somewhat surprisingly, the quantitative values for the thermodynamic frequency uncertainty limits given in the current literature are not consistent with each other, illustrating the lack of a simple, general and consistent approach for calculating it.

Here we provide an intuitive and general approach to quantify the thermodynamic limit of the resonance frequency measurement. We derive the Cramer-Rao lower bound for the statistical uncertainty of the frequency measurement for a classical linear harmonic oscillator subject to the thermodynamic Langevin force under continuous position detection. We also present a straightforward averaging formula for calculating the maximum-likelihood resonance frequency from a time series of oscillator position measurements. We use numerically simulated fluctuating oscillator position data to show that the frequency estimator is statistically efficient, i.e. the resulting frequency statistical uncertainty reaches the CRLB for averaging times both above and below the relaxation time. Remarkably, a continuous root-mean-square dependence on averaging time is predicted and observed down to times below the relaxation time for fluctuating oscillators, as long as within the measurement bandwidth the thermal force noise is white and integrated position readout noise power is negligible compared to thermal fluctuation power.

We explicitly limit this analysis to the resonator without external drive and with negligible readout noise, while we intend to extend the formalism to the driven resonators and the noisy detection in follow-up work. Although the frequency measurement uncertainty generally improves when the resonator is driven, the no-drive limit is important to understand. First, the resonator thermal fluctuations without drive are commonly recorded in many modern micro- and nanoscale systems experiments. Additionally, as precision nanophotonic readouts become available in chip-scale sensors, this fluctuation-based, passive mode of mechanical frequency sensing may find practical applications. It simplifies the sensor while providing a wide dynamic range beyond the resonator bandwidth. Such sensors would also require the real-time, dynamic frequency estimator working down to short averaging times and at fundamental precision limits.

The equation of motion for a harmonic resonator subject to thermodynamic noise is written as:

$$\ddot{x} + \Gamma \dot{x} + \omega_0^2 x = \frac{f}{m} \qquad (1)$$

where $x$ is the position of the resonator, $\Gamma$ is the damping factor, $\omega_0$ is the resonance frequency, $m$ is the effective mass of the mode, and $f$ is the stochastic Langevin force. From Boltzmann distribution and equipartition theorem, in thermal equilibrium, the position $x$ follows zero-mean Gaussian distribution with variance $\sigma^2$ given by $\langle x^2 \rangle \equiv \sigma^2 = \frac{k_b T}{m \omega_0^2}$, where $k_b$ is the Boltzmann constant and $T$ is the effective temperature.

By defining a slowly varying variable $u$ via $x = \frac{1}{2}(u e^{i\omega t} + u^* e^{-i\omega t})$, we rewrite the equation of motion in the rotating wave approximation (RWA):

$$\dot{u} + \frac{\Gamma}{2} u + i \Delta \omega \, u = \frac{f_1 - i f_2}{i \omega m} \qquad (2)$$

where $\Delta \omega = (\omega - \omega_0) \ll \omega_0$ and $f_{1,2}$ are the in-phase and quadrature components of the Langevin force near resonance.

Experimentally $u = X + iY$ can be directly measured by a homodyne detector such as a lock-in amplifier with a local oscillator frequency $\omega$ and a sufficiently high bandwidth $\gg \Gamma$.

From $\langle x^2 \rangle = \frac{1}{4}\langle (ue^{i\omega t_0} + u^*e^{-i\omega t_0})^2 \rangle = \frac{1}{4}\langle 2uu^* \rangle$, we obtain:

$$\langle |u|^2 \rangle = 2\sigma^2 \tag{3}$$

In thermal equilibrium, $u$ obeys a zero-mean two-dimensional Gaussian distribution with a variance of $\sigma^2$ for both in-phase and quadrature components.

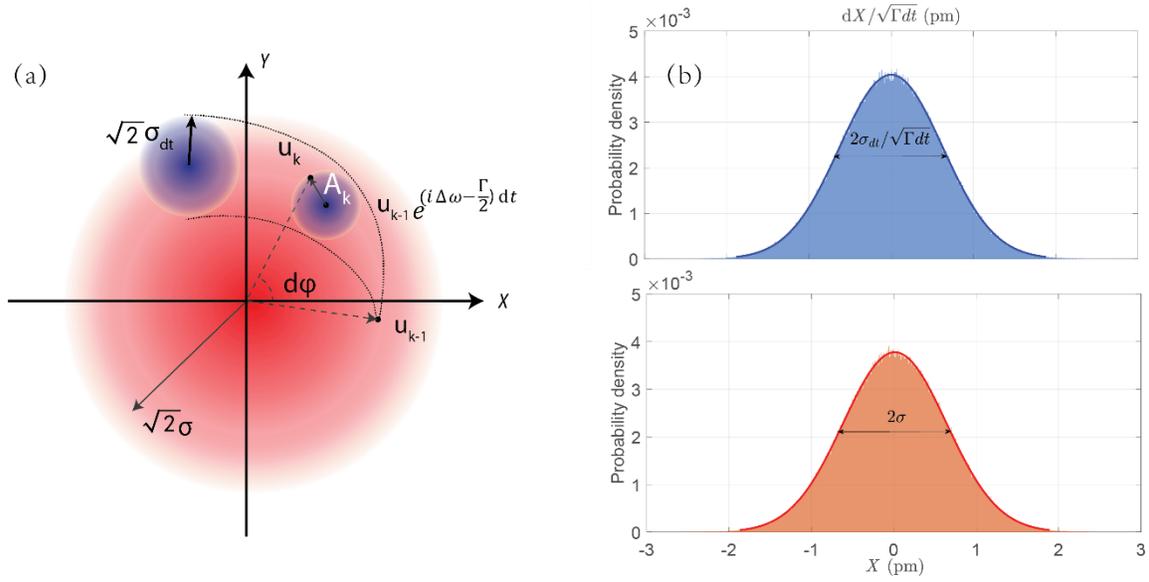

Figure 1 Thermal fluctuation induced phase diffusion. (a) Schematic of phase diffusion. (b) Probability density of $X$ (lower) and $\frac{dX}{\sqrt{\Gamma dt}}$ (upper) of simulated results, on the same $X$-axis.

In the continuous detection limit, a series of values $u_j$ is measured at time intervals $dt \ll 1/\Gamma$. Following Eq. (2), the resonator rotates around origin at the rate $\Delta\omega$, decays at the rate $\frac{\Gamma}{2}$ and diffuses in response to the Langevin force, going from a position $u_{k-1}$ to the next position $u_k$ in time $dt$, as shown in the $X$-$Y$ phase diagram in Figure 1(a). Given a known value of $u_{k-1}$, the probability distribution of $u_k$ in the phase diagram is a 2-dimensional Gaussian (small blue bubble) with a mean value of $u_{k-1}e^{(i\Delta\omega - \frac{\Gamma}{2})dt}$ and a small variance $\sigma_{dt}^2$ for each dimension:

$$p(u_k|u_{k-1}) = \frac{1}{2\pi\sigma_{dt}^2} e^{-\frac{\left|u_k - u_{k-1} e^{(i\Delta\omega - \frac{\Gamma}{2})dt}\right|^2}{2\sigma_{dt}^2}} \tag{4}$$

The evolving step size and thermal uncertainty become larger with increasing measurement time interval (big blue bubble). If the time interval $dt > 1/\Gamma$, $u_k$ does not correlate to $u_{k-1}$ anymore. In the following $dt \ll 1/\Gamma$ is assumed.

The variance $\sigma_{dt}^2$ can be related to $\sigma^2$ by noting that in thermal equilibrium the decay and thermal fluctuations balance each other, resulting in a steady-state. From Eq. (4) it follows that:

$$\langle|u_k|^2\rangle = \langle|u_{k-1}|^2\rangle e^{-\Gamma dt} + 2\sigma_{dt}^2 \tag{5}$$

where the average is over all pairs $(k-1, k)$ in the equilibrium ensemble.

In steady-state, from Eq. (3) $\langle|u_k|^2\rangle = \langle|u_{k-1}|^2\rangle = 2\sigma^2$ and for $\Gamma dt \ll 1$:

$$\sigma_{dt}^2 = \Gamma dt \sigma^2 \tag{6}$$

To numerically model the resonator described by Eq. (1), Eq. (2), we note that in Eq. (1) the Langevin force satisfies $\langle f(t)f(t')\rangle = 2\Gamma k_b T m \delta(t-t')$ and therefore on each given short time interval $dt$ the $f(t)$ can be modeled as having a random value picked from a zero-mean Gaussian distribution with a variance $\frac{2\Gamma k_b T m}{dt}$ [24], [25]. In RWA the values for $f_{1,2}$ are each picked from a zero-mean Gaussian with the variance $Var(f_1) = Var(f_2) = \frac{\Gamma k_b T m}{dt}$.

Figure 1(b) shows the distribution of $X$ and $\frac{dX}{\sqrt{\Gamma dt}}$ of simulated results from Eq. (2) with experimentally realistic parameters [19]: $m$ = 1 pg, $T$ = 300 K, $\omega_0/2\pi$ = 27.76 MHz, $\Gamma/2\pi = 620$ Hz, $\Delta\omega/2\pi = 0$ Hz and $dt = 10$ μs. The corresponding standard deviation for 10 s of simulated data are $\sigma = (625.10 \pm 1.54)$ fm and $\sigma_{dt}/\sqrt{\Gamma dt} = (626.08 \pm 2.08)$ fm, respectively, which is consistent with $\sigma = \sqrt{\frac{k_b T}{m\omega_0^2}} = 624.67$ fm. All uncertainties are one standard deviation statistical uncertainties, unless otherwise noted.

The theoretical thermodynamic frequency detection limit is calculated through its Cramer-Rao lower bound [26], [27]:

$$\text{Var}(\Delta\omega) \geq -\left[\text{E}\left(\frac{\partial^2}{\partial\Delta\omega^2}\ln P(U,\Delta\omega)\right)\right]^{-1} \tag{7}$$

by considering the 2N dimensional probability density $P(U,\Delta\omega)$ of obtaining a specific series of $U = \{u_1 \ldots u_N\}$ from $N$ measurements. E denotes the expectation for a given $\Delta\omega$. The probability density is:

$$P(U,\Delta\omega) = \prod_{k=1}^{N} p(u_k,\Delta\omega) = \frac{1}{2\pi\sigma^2} e^{-\frac{|u_1|^2}{2\sigma^2}} \prod_{k=2}^{N} p(u_k|u_{k-1}) \tag{8}$$

Without any prior knowledge, the first position $u_1$ obeys two-dimensional zero-mean Gaussian distribution with a variance of $\sigma^2$ in each dimension. After knowing the first position $u_1$, the probability of latter positions $u_k$ is obtained from the recursive formula given in Eq. (4).

The natural logarithm of the probability density is:

$$\ln P(U,\Delta\omega) = C - \frac{|u_1|^2}{2\sigma^2} - \sum_{k=2}^{N} \frac{\left|u_k - u_{k-1}e^{\left(i\Delta\omega-\frac{\Gamma}{2}\right)dt}\right|^2}{2\sigma_{dt}^2} \tag{9}$$

where $C$ is a parameter independent from $\Delta\omega$.

We define $A_k \equiv u_k - u_{k-1}e^{\left(i\Delta\omega-\frac{\Gamma}{2}\right)dt}$, which obeys 2-dimensional zero-mean Gaussian distribution with a variance of $\sigma_{dt}^2$ for each dimension, given by the probability in Eq. (4), i.e. $\langle A_k \rangle = 0$, $\langle |A_k|^2 \rangle = 2\sigma_{dt}^2$. $A_k$ describes a $k$th diffusion step and is independent and therefore uncorrelated to $u_{k-1}$, i.e. $\langle A_k u_{k-1}^* \rangle = 0$. We further define $B_k \equiv A_k' \equiv \frac{\partial A_k}{\partial\Delta\omega} = -idt u_{k-1}e^{\left(i\Delta\omega-\frac{\Gamma}{2}\right)dt}$ and $B_k' \equiv \frac{\partial B_k}{\partial\Delta\omega}$. Note that $\langle B_k' A_k^* \rangle = dt^2 e^{\left(i\Delta\omega-\frac{\Gamma}{2}\right)dt} \langle u_{k-1} A_k^* \rangle = 0$.

Taking the second derivative of Eq. (9), we obtain:

$$-\langle\frac{\partial^2}{\partial\Delta\omega^2}\ln P(U,\Delta\omega)\rangle_\tau = -\langle\sum_{k=2}^{N} \frac{B_k'A_k^* + B_k B_k^* + c.c.}{2\sigma_{dt}^2}\rangle_\tau = -\langle\sum_{k=2}^{N} \frac{B_k B_k^* + c.c.}{2\sigma_{dt}^2}\rangle_\tau =$$

$$(N-1)\frac{dt^2 e^{-\Gamma dt}}{2\sigma_{dt}^2}\langle u_{k-1}u_{k-1}^* + c.c.\rangle_\tau = (N-1)\frac{dt^2 e^{-\Gamma dt}}{2\sigma_{dt}^2}4\sigma^2 = \frac{2\tau}{\Gamma}e^{-\Gamma dt} \tag{10}$$

where $\tau = (N-1)dt$ is measurement time. Based on Eq. (7) we obtain that the Cramer-Rao lower bound as:

$$\text{Var}(\Delta\omega) \geq \frac{\Gamma}{2\tau} e^{\Gamma dt} \approx \frac{\Gamma}{2\tau} \qquad (11)$$

We now derive the frequency estimator for a measurement $U$ that returns the most likely $\Delta\omega$, which satisfies $\partial P(U, \Delta\omega)/\partial \Delta\omega = 0$, or, equivalently, $\frac{\partial}{\partial \Delta\omega} \ln P(U, \Delta\omega) = 0$. Simplifying Eq. (9) in the continuous detection limit $\Gamma dt \ll 1$, $e^{(i\Delta\omega - \frac{\Gamma}{2})dt} \approx 1 + i dt \left(\Delta\omega + i\frac{\Gamma}{2}\right)$, we obtain:

$$\Delta\omega = \frac{\sum_k [(iu_k \dot{u}_k^* - i u_k^* \dot{u}_k)]}{2\sum_k u_k u_k^*} \qquad (12)$$

where we have defined $\dot{u}_k \equiv \frac{u_k - u_{k-1}}{dt}$.

This expression can be intuitively understood in polar coordinates, where $u_k = |u_k| e^{i\varphi_k}$ and $\dot{\varphi}_k$ is defined via $\dot{u}_k = |\dot{u}_k| e^{i\varphi_k} + i|u_k| e^{i\varphi_k} \dot{\varphi}_k$. Eq. (12) can be rewritten as $\Delta\omega = \frac{\sum_k |u_k|^2 \dot{\varphi}_k}{\sum_k |u_k|^2}$, which is an average of frequency estimates $\dot{\varphi}_k$ for each step, weighted by $|u_k|^2$. The measured phases have smaller uncertainties when the amplitudes are larger. For a small enough $dt$, $\sigma_{dt} \ll |u_k|$ for almost all samples, and $\text{Var}(\dot{\varphi}_k) = \frac{\sigma_{dt}^2}{(dt |u_k|)^2}$ since $\dot{\varphi}_k = \frac{\varphi_k - \varphi_{k-1}}{dt} \approx \frac{1}{dt} \frac{(u_k - u_{k-1})_{azimuthal}}{|u_k|}$, $(u_k - u_{k-1})_{azimuthal}$ has a variance of $\sigma_{dt}^2$ and $|u_k|$ is approximately a constant for a short period of time (Figure 1a). The weight factors $|u_k|^2$ are proportional to the inverse variances, and the estimator given in Eq. (12) is a conventional, error-weighted average of $\dot{\varphi}_k$.

Practically, the estimator can be very simply implemented as a running average and is fast computationally, scaling linearly with the number of samples and requiring no Fourier transforms or iterative procedures.

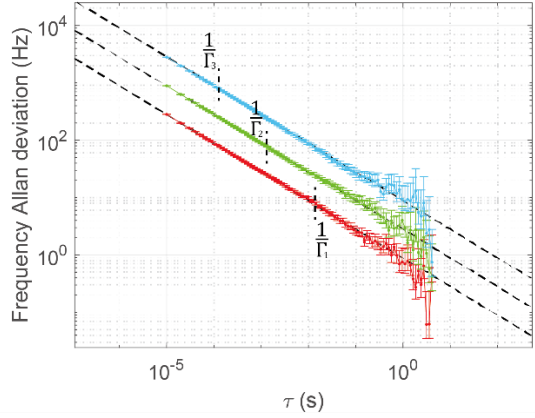

Figure 2 Frequency Allan deviation of estimated frequency and the corresponding Cramer-Rao lower bound. Red, green and blue data are for $\Gamma_{1,2,3}/2\pi$ = 10 Hz, 100 Hz and 1 kHz, respectively. Short black segments label the relaxation time $\tau = 1/\Gamma_{1,2,3}$, respectively. The uncertainties for Allan deviation are determined by Chi-Squared Confidence Intervals.

To verify that our estimator is statistically efficient, i.e. achieves frequency uncertainty at the CRLB, we apply it to estimate frequency from simulated motion data and calculate the weighted Allan variance as:

$$\sigma_f^2(\tau) = \frac{1}{2} \langle W_{k\tau} [\Delta\omega_{(k+1)\tau}/2\pi - \Delta\omega_{k\tau}/2\pi]^2 \rangle_{T_0} \qquad (13)$$

where $\Delta\omega_{k\tau}$ represents the frequency estimated from the data in a time interval $[(k-1)\tau, k\tau]$ and $\langle ... \rangle_{T_0}$ represents the average of the data over the full-time trace of length $T_0$. The weights are $W_{k\tau} = \frac{\langle |u|^2 \rangle_{k\tau}}{\langle |u|^2 \rangle_{T_0}}$ and tend to conventional unity weights for the $\tau > 1/\Gamma$, while deviating from unity at small $\tau$. Although there is no significant difference between weighted and unweighted Allan deviation for $\tau > 1/\Gamma$, for short time scales the widely used unweighted Allan deviation is only appropriate for driven resonators, where $W_{k\tau} \approx 1$ on all time scales.

The data points of Figure 2 show the Allan deviation $\sigma_f(\tau)$ of frequency estimated from the simulated motion data with the parameters same as those used in Figure 1 (b). Different colors correspond to $\Gamma/2\pi$ = 10 Hz, 100 Hz and 1 kHz. The dashed lines are CRLB, calculated by Eq. (11) without adjustable parameters. The good agreement between the weighted Allan deviation of our estimated frequency and the CRLB proves that our frequency estimator is efficient. In the absence of readout noise, the estimator performs at CRLB even for the averaging times below the relaxation time $1/\Gamma$ of the resonator, as indicated on Figure 2. Since the resonator motion amplitude fluctuates on $\approx 1/\Gamma$ timescale, so does the uncertainty of the frequency measurements. Therefore, it is important to note that for $\tau < 1/\Gamma$ the CRLB and the Allan deviation indicate the frequency uncertainty averaged over many repeated short measurements, collectively spanning $T_0 \gg 1/\Gamma$.

In conclusion, we derived a Cramer-Rao lower bound (11) for the resonance frequency uncertainty for a linear resonator subject to thermal fluctuations. We present an easy-to-compute maximum-likelihood estimator (12) for resonance frequency from motion data and use numerically simulated motion data to show that the Allan deviation of the estimated frequency reaches the CRLB uncertainty. The results are valid on time scales both above and below the resonator relaxation time, specifically addressing frequency detection in the limit of the resonator driven by thermal fluctuations alone. The CRLB approach presented here is quite general and may be fruitfully extended to many other systems by analyzing the frequency dependence of the corresponding measurement data vector probability density. Beyond direct extension to harmonically driven linear resonators in the presence of measurement uncertainty, the approach may prove useful for understanding more complex driven oscillators. These include oscillators in the nonlinear regime exhibiting amplitude-phase-noise and mode mixing, which have been exploited to achieve better frequency stabilization [5], [6], [28].